\documentclass[twocolumn,aps,prb,floatfix,showpacs]{revtex4}
\usepackage{graphics}
\usepackage{epsfig}
\usepackage{bm}
\begin{document}

\title{Bosons in fluctuating gauge fields: Bose
metal and phase separation 
}

\author{Robert L. Jack}
\author{Derek K. K. Lee}
\affiliation{Blackett Laboratory, Imperial College, Prince Consort Road,
 London SW7 2BW, United Kingdom}

\begin{abstract}
  We study a two-dimensional system of bosons interacting with a
  fluctuating $U(1)$ gauge field with overdamped dynamics. We find two
  instabilities of the condensed phase at $T=0$: one to phase
  separation and another to a homogeneous non-superfluid (Bose
  metal). The presence of both instabilities in the model is dependent
  on the low-energy form of the gauge field propagator. We discuss the
  relevance of our findings to the U(1) gauge theory of the $t$--$J$
  model.
\end{abstract}
%PACS: 74.20.Mn :Theories and models of superconducting state  
%            - nonconventional mechanisms (inc RVB)
%      67.40.Db : Boson degeneracy and superfluidity of 4He
%            - Quantum statistical theory; ground state, elementary excitations
\maketitle

\section{Introduction}

A Bose liquid naturally becomes a superfluid at low temperatures.  The
existence of a non-superfluid liquid state (the ``Bose metal'') down
to zero temperature has been a subject of recent
controversy~\cite{DasDoniach,DalidRG,DalidInteraction}. Such a
scenario might arise if the bosons interact with fluctuating external
fields~\cite{DasDoniach,IoffeLarkin,LeeNag,Ioffe,LeeLeeKim}. A
fluctuating vector potential couples directly to the gradient of the
phase of the superfluid order parameter. This makes it a natural
choice for destroying long-range phase coherence in a superfluid.

In this paper, we study how a vector potential with overdamped
dynamics might destroy superfluidity.  Our choice of overdamped
dynamics is motivated by the gauge theory~\cite{IoffeLarkin,LeeNag} of
the $t$--$J$ model, relevant to the cuprate superconductors.  The
existence of a Bose metal far below the degeneracy temperature is
crucial for the picture of spin-charge separation in that theory.

Our study is the first to consider the damping of the zero-point
fluctuations of the gauge field. We find that a metallic phase may
exist in the part of the phase diagram relevant to the cuprates.
Related models with different dynamics have been studied before. Using
a quasistatic approximation for the gauge field, Lee, Lee and
Kim~\cite{LeeLeeKim} showed the loss of superfluidity in a quantum
Monte Carlo simulation. On the other hand, a metallic phase was not
found with a propagating (``Maxwell'') gauge
field~\cite{Ioffe}. Instead, the system is unstable to phase
separation (unless it is stabilised by long-ranged repulsion.)

In our treatment, we use the perturbative corrections
calculated by Feigelman \emph{et al}~\cite{Ioffe} in a simple
renormalisation scheme. In Ref. \onlinecite{Ioffe}, a self-consistent scheme 
was used to calculate the superfluid density in the presence of
fluctuating magnetic fields. It was clear from that work that the
superfluid response depends on the compressibility of the Bose liquid,
and \emph{vice versa}: the superfluid expels flux over a penetration
depth, but that inevitably reduces the local density, exciting phonons
and causing an apparent increase in the compressibility. Since there
is the possibility of phase separation (crudely speaking, the bosons
and magnetic flux expelling each other), we want to track this
inter-dependence in more detail. Our scheme resembles a
frequency-dependent mean-field theory. We
successively integrate out high-frequency fluctuations, generating at
each stage corrections to the low-energy response functions --- the
superfluid density $n_s$ and the compressibility $\kappa$. We believe
that this will give a better picture of the instabilities of the
superfluid phase.

The form of the paper is as follows. We start by reviewing the method
by which charge dynamics of the $t$--$J$ model can be obtained from a
model of bosons interacting with a $U(1)$ gauge field. We then calculate
corrections to the compressibility, phase stiffness and Meissner response,
in a perturbation expansion about the superfluid state. We use these 
corrections in 
a renormalisation scheme, which we apply to systems with both overdamped and
propagating gauge fields. Finally, we discuss our results and make some 
conclusions about the effect of different gauge fields on the bosonic systems.

\section{The model}
\label{sec:tJ}

We will now describe in more detail our model, in particular, its
motivation from the gauge theory of the $t$--$J$ model. 
In this paper, we focus on two-dimensional bosonic systems with
(imaginary-time) Lagrangian densities of the form: ($\hbar=c=e=1$)
\begin{eqnarray}
\label{equ:L_psi}
\mathcal{L}_{x} & = & b_x^\dagger ( \partial_\tau - \mu ) b_x 
  + b_x^\dagger {1\over 2m} (i \nabla - {a}_x)^2  b_x + 
\nonumber \\ 
& &  \frac{1}{2} U (b_x^\dagger b_x)^2 + 
\frac{1}{2} \int \mathrm{d}x' \, {a}_x^\mu (D^{\mu\nu}_{x-x'})^{-1} 
  {a}^\nu_{x'} 
\end{eqnarray} 
where $b_x$ is a complex bosonic field with local repulsion $U$ and
mass $m$, and ${a}_x^\mu$ is a (two-vector) gauge field with
propagator $D$. We label both spatial and temporal coordinates by
$x=(\bm{r},\tau)$. 

As already mentioned, we study the gauge fields that arise from the
U(1) gauge theory of the $t$--$J$ model. The $t$--$J$ model describes
electrons with hopping integral $t$ and antiferromagnetic exchange
$J$. The Coulomb repulsion is so strong that we assume that there is
at most one electron per site. The Hamiltonian is:
\begin{equation}
\label{equ:tJ}
H = -t \sum_{\langle ij \rangle} \left( c_{i\sigma}^\dagger c_{j\sigma}
  + \hbox{h.c.} \right)
     + J \sum_{\langle ij \rangle} \left( {\bf S}_i \cdot {\bf S}_j
     - \frac{n_i n_j}{4} \right)
\end{equation}
with a constraint of no double occupancy at each site. The operator
$c^\dagger_{i\sigma}$ creates an electron at site $i$ with spin
$\sigma$, $n_i$ is the number of electrons on a site, and ${\bf S}_i =
\sum_{\alpha \beta} c_{i\alpha} \bm{\sigma}_{\alpha \beta}
c_{i\beta}$ is the spin operator at a site. The sums are taken over
pairs of nearest neighbours on a square lattice. 

A ``slave boson'' treatment~\cite{Coleman,ReadNewns} of the
constraint of single occupancy reduces this model to a Lagrangian density
of the form in equation~\ref{equ:L_psi}.  The electronic operator is
split into a spin-half fermion and a charge-$e$ boson:
$c_{i\sigma}=b^\dag_i f_{i\sigma}$.  The single-occupancy constraint
means the fermion and boson currents must be equal and opposite:
$J_\mathrm{f}=-J_b$.  The gauge theory enforces this constraint by
making the boson and fermion currents interact indirectly \emph{via} a
vector potential. 

After this (exact) decoupling, 
we may obtain an effective theory for the bosons. The derivation of
this (approximate) theory is shown in appendix~\ref{sec:app-tj}. 
The effective Lagrangian density for the bosons takes the form of 
equation~\ref{equ:L_psi}; the theory aims to capture the strong 
correlations in the $t-J$ model, 
which arise from the constraint of no double occupancy.

As explained in appendix~\ref{sec:app-tj},
the scalar part of the gauge field that arises in this treatment has been
absorbed into the local repulsion $U$.
The boson density is the charge doping away from half-filling. This
appears to be consistent with the charge carrier density from
transport measurements. So, we will concentrate on boson densities
small compared to integer filling of the lattice and we can ignore the
possibility of a Bose Mott insulator in our system.

The excitations of the fermionic subsystem can exchange energy and
momentum with the bosons. Integrating out the fermions amounts to
treating them as an effective medium through which the gauge field
propagates. In fact, the Lagrangian for the gauge field (last term in
(\ref{equ:L_psi})) is determined by the dynamics of the particle-hole
excitations of the fermions. It can be shown~\cite{IoffeLarkin} that
the propagator, $D_{\omega_n,q}$, is directly related to the fermion current
response functions. This is consistent with the constraint
on the boson and fermion currents. 

Let us choose the gauge $\nabla \cdot \bm{a}=0$ so that only the
transverse component, $a^\perp$, is non-zero and $D^{\mu\nu}$
has only one degree of freedom which we call $D$.
As shown in appendix~\ref{sec:app-tj}, the 
spectral density of this propagator is $A(\omega,q)= 2\, \mathrm{Im}
[ - 1/\Pi^T_\mathrm{ff}(\omega,q) ]$ where
$\Pi^T_\mathrm{ff} = (\delta_{\mu\nu} - {q_\mu q_\nu / q^2})
[ i \langle {J_\mathrm{f}^\mu} J_\mathrm{f}^\nu
\rangle^R_{\omega,q} - (n_\mathrm{f}/
m_\mathrm{f})\delta^{\mu\nu} ]$ is the retarded transverse
polarisation of the fermions.

For simplicity, we will model the fermions as an isotropic Fermi
liquid with mass $m_{\rm f}$ and a density $n_{\rm f}$ which we choose
to be close to the half-filling density on a lattice model. For the
cuprates, this is expected to be reasonable near optimal doping. In
other words, we ignore the ``pseudogap'' spin physics in underdoped
materials. 

With this choice of isotropic fermions, the polarisation can be
derived analytically. This saves a great deal of computational time in the 
numerical calculations carried out in the next section.
However, it ignores the underlying lattice. 
For this reason we will introduce a high-energy cutoff, $\Omega$, in the 
spectral density of the gauge field propagator. 
We choose $\Omega=E_\mathrm{F}$ so that particle-hole
excitations with energies greater than the Fermi energy $E_\mathrm{F}$
are excluded --- these would have involved quasiparticles outside the
Brillouin zone of the underlying lattice. The form of the polarisation 
and the effect of the cutoff $\Omega$ are discussed in appendix~\ref{app:Pi}.
Taking all this into account, we obtain:
%
%overall - sign
\begin{equation}
\label{equ:D_lindhard}
D_{\omega_n,q} = \int_{-\Omega}^\Omega {\mathrm{d} \omega' \over 2 \pi}
  {1\over \omega' - i \omega_n}
\left( - 2 \mathrm{Im} \left[ 
   { 1\over \Pi_\mathrm{ff}^T(\omega',q) } 
  \right] \right)
\end{equation}
At small frequencies and momenta this reduces to the form
$D_{\omega_n,q} = \left( \sigma_q |\omega_n| + \chi q^2 \right)^{-1}$
where $\chi=(12 \pi m_\mathrm{f})^{-1}$ is the diamagnetic
susceptibility of a two-dimensional Fermi liquid and
$\sigma_q={2n_\mathrm{f}/ k_\mathrm{F} q}$ is the conductivity due to
Landau damping. (However, we do not work with this simple form
because we find that strong scattering occurs at large wavevectors.)

Equation~\ref{equ:D_lindhard} completes our definition of the model in 
(\ref{equ:L_psi}) by specifying the spectrum of gauge field fluctuations.
In the following sections
we discuss instabilities of the superfluid state in this model.

\section{The renormalisation scheme}

In this section, we investigate instabilities of the superfluid state in
the general model given by equation~\ref{equ:L_psi}. We discuss the 
dependence of these instabilities on the form of the gauge field propagator.
The interaction between
the bosonic excitations and the gauge field affects the compressibility of 
the bosons and their phase stiffness, as well as the Meissner response of
the gauge field. These corrections can be calculated in perturbation theory,
but near instabilities of the superfluid state a more advanced treatment is
necessary, so we introduce a simple renormalisation scheme. 

We first expand about a uniform condensate by writing: $\Psi_x =
\sqrt{n+\rho_x} e^{i\theta_x}$, where $n$ is the mean boson density,
$\rho$ represents density fluctuations, and $\theta$ represents
fluctuations in the phase of the bosonic field.  
The Lagrangian density becomes 
$\mathcal{L}_x  = \mathcal{L}^0_x + \mathcal{L}^\mathrm{int}_x$,
where
\begin{eqnarray}
\mathcal{L}^0_x & = & i \rho_x \partial_\tau \theta_x 
+    \frac{U}{2}\rho_x^2  
  + {1\over 8mn} \left|\nabla \rho_x \right|^2 
        \nonumber \\ & &
  + {n \over 2m} \left|\nabla \theta_x \right|^2 
+ {1\over 2} \int\!\! \mathrm{d}x' \,  a^\perp_x ( D^0_{x-x'} )^{-1}
  a^\perp_{x'} 
\label{equ:L_rho_theta}
\end{eqnarray}
and
% 
% $a$ not bold
\begin{eqnarray}
\mathcal{L}^\mathrm{int}_x & = & 
  {1\over 2m} (\nabla\theta_x + a_x^\perp)^2 \rho_x - 
  {\rho_x\over 8mn(n+\rho_x) } |\nabla\rho_x|^2
\label{equ:L_int}
\end{eqnarray}
The propagator $( D^0_{x-x'} )^{-1} = 
\left( {n\over m}\delta (x-x') + D^{-1}_{x-x'} \right)$ where the first
term is the Meissner response which modifies the bare propagator.

The Lagrangian density $\mathcal{L}^0$ describes a superfluid Bose
liquid with compressibility $\kappa = U^{-1}$ and a superfluid density
$n_s=n$.  We are interested in how the interaction terms in
$\mathcal{L}^\mathrm{int}$ affect these quantities. The system will remain
superfluid as long as both of these quantities are finite.

Setting $a=0$ in the action above corresponds to the neutral superfluid. The
resulting action is not quadratic, but neither of the remaining interaction 
terms modify the superfluid density (this must be equal to the total density
for a Galilean invariant system). We therefore drop these terms, keeping only
interaction terms that contain the gauge field, $a$.
If the discarded terms have an effect on the compressibility then this can be 
absorbed into $U$. 

Any effect on the superfluid response must therefore come from the terms
involving the gauge field. Simple perturbation theory is not suitable for 
discussing the instabilities at small $n_s$ and small $U$, since strong 
renormalisation of these quantities  renders it
invalid. Feigelman~\emph{et~al.}~\cite{Ioffe} used a self-consistent
method to discuss the reduction of the superfluid response in a system
with long-ranged repulsion (so the compressibility is not
renormalised).

\begin{figure}
\epsfig{file=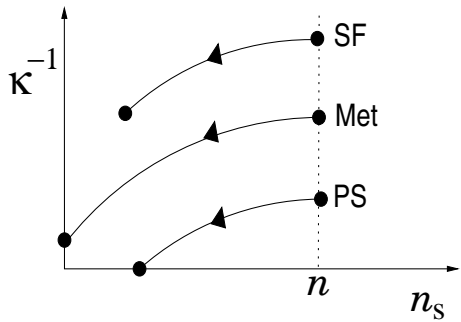, width=0.7\columnwidth}
\caption{Possible renormalisation flows. 
  The superfluid (SF) is stable for a large initial value of the
  inverse compressibility $\kappa^{-1}$ ($=U$). A small initial $\kappa^{-1}$
  may lead to phase separation (PS), while intermediate
  values give rise to a metal, especially at small doping.  }
\label{fig:flows}
\end{figure}

We use a different method, which allows us to treat the corrections to
both the gauge field propagator and the bosonic propagator on an equal 
basis. We integrate out the highest frequency fields, which gives an
effective theory for the remaining fields, with renormalised
superfluid fraction and compressibility. This process is then repeated
until only the very smallest frequencies are left.

As the bandwidth is reduced, three scenarios are possible; these are shown
schematically in figure~\ref{fig:flows}. If both $n_s$ and $\kappa$ remain
finite after all the fields have been integrated then the system will remain
superfluid. If $\kappa$ diverges at finite $n_s$ then there will be an
instability to phase separation, but if $n_s$ vanishes at finite $\kappa$ then
there will be an instability to a homogeneous non-superfluid, which we call
a metal in the rest of this paper.

\begin{figure}
\epsfig{file=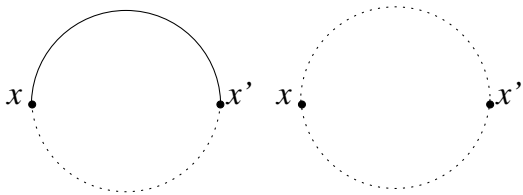, width=0.8\columnwidth}
\caption{Polarisation insertions used in one-loop perturbational
calculations. Dotted line is a gauge field propagator, solid line
is a propagator for phonons in the Bose liquid. The processes affect
the phase stiffness and Meissner response (left) and the compressibility 
(right).}
\label{fig:pt}
\end{figure}

To proceed, we generalise our model to allow for a superfluid density 
that is not equal to the mean boson density and a compressibility that is 
not equal to $U^{-1}$. We write
$\mathcal{L}_x  = \mathcal{L}_x^s + \mathcal{L}_x^1$,
where
%
% transverse label on $a$
\begin{eqnarray}
\mathcal{L}^s_x &=&  i \rho_x \partial_\tau \theta_x 
+    {1\over 2\kappa} \rho_x^2 + {1\over 8mn} 
  \left| \nabla \rho_x \right|^2 
\nonumber \\ &+&
  {n_s \over 2m} \left| \nabla \theta_x \right|^2 
+  {1\over 2} a_x^\perp \left( D^s_{x-x'} \right)^{-1} a_{x'}^\perp 
\end{eqnarray}
and 
\begin{eqnarray}
\mathcal{L}^1_x & = & 
  {1\over m} \rho_x (\bm{a}_x  \cdot \nabla \theta_x) 
  + {1\over 2m} (a_x^\perp)^2 \rho_x
\end{eqnarray}
with $(D^s_{x-x'})^{-1}= (n_s / m) \delta( x-x') + D^{-1} $. Note
that the phase stiffness of the bosons and the Meissner response of
the gauge field are characterised by the same $n_s$. This is
required by the underlying U(1) gauge symmetry. In the perturbational
calculation the correction to the phase stiffness comes from the first
term in $\mathcal{L}_1$ and the correction to the Meissner response
comes from the second term. However, both corrections are given by the
same polarisation insertion (the first bubble in figure~\ref{fig:pt});
this ensures that they are equal, as required.

We start the renormalisation process with $\kappa^{-1}=U$ and
$n_s=n$. At each stage, we integrate out all boson excitations with
frequencies whose magnitudes lie between the maximum $\omega_n$ and
$\omega_n-\delta \omega$. We then average over fast fluctuations in
the gauge field, as done in Refs. \onlinecite{Ioffe,DasDoniach}. As each
frequency shell is integrated out, we correct the values of $n_s$ and
$\kappa$ and use these values, denoted as $n_s(\omega_n)$ and
$\kappa(\omega_n)$, for the next shell. 

The one-loop corrections to $n_s$ and $\kappa$ are given by:
\begin{equation}
{\delta n_s \over n_s} = - (\delta \omega)  \Pi^{(1)}_{\omega_n} \; ,
\qquad\delta (\kappa^{-1}) = - (\delta \omega) K^{(1)}_{\omega_n}
\end{equation}
where
\begin{equation}
\label{equ:K1_scaling}
K^{(1)}_{\omega_n} =  
{1\over 2 \pi m^2} \int_0^\infty 
{ q \mathrm{d} q \over 2\pi} 
\left( {D}_{\omega_n,q}^s \right)^2
\end{equation}
and
\begin{eqnarray}
%\lefteqn{\hspace{-15pt} 
\Pi^{(1)}_{\omega_n} =
\frac{1}{\pi m^2}  
\int_0^\infty\frac{q \mathrm{d} q}{2\pi}
{D}_{\omega_n,q}^s {C}_{\omega_n,q}
\label{equ:Pi1_scaling}
\end{eqnarray}
These depend on the gauge field propagator $D$ (about which we have so far
assumed nothing) and the propagator for phonons in the Bose liquid:
\begin{eqnarray}
{C}_{\omega_n,q}  &=&  
\frac{m}{2 n_s} \langle
  \rho_{\omega_n,q} \rho_{-\omega_n,-q}\rangle
\nonumber\\
&=& {q^2 \over \omega_n^2 + n_s q^2 ( {\kappa^{-1} \over m} +
{ q^2 \over 4 m^2 n} ) }
\end{eqnarray}

These corrections correspond to the diagrams in figure~\ref{fig:pt}. 
To reiterate, the renormalisation scheme is implemented by using the
renormalised values of $n_s$ and $\kappa$ when evaluating
$C_{\omega_n}$ and $D_{\omega_n}$ at each stage of the procedure.
Perturbation theory
corresponds to performing the frequency integrals keeping $n_s$ and $\kappa$
constant. Our renormalisation scheme resembles a frequency dependent 
mean-field theory in that these parameters are constant with respect
to momentum, but dependent on frequency.
We see that $\delta n_s$ is negative and $\delta \kappa$ is positive: the 
perturbations reduce the superfluid correlations and increase the density
fluctuations.

\subsection{Overdamped propagator}

We now carry out this renormalisation scheme using the gauge field propagator 
given in equation~\ref{equ:D_lindhard} 
and investigate the resulting phase diagram.
Let us now identify four independent dimensionless parameters for the
system. We choose $\lambda_{0}^2=(n_\mathrm{f}/n)(m/m_\mathrm{f})$,
$U_r=m U$, $m_r=m_\mathrm{f}/ m$ and $\Omega_r=\Omega/E_\mathrm{F}$.
$\lambda_{0}$ is the penetration depth of the Bose superfluid at
$n_s=n$ in units of the fermion separation (\emph{i.e.}  lattice
spacing at $n_\mathrm{f}=1$). With the other dimensionless parameters
fixed, varying $\lambda_0$ corresponds to changing the boson density
$n$: $\lambda_0\propto 1/n^{1/2}$. $U_r$ measures the relative size of
the repulsive and kinetic energies of the bosons. $U_r$ should be at
least O(1) if we want to model the hard-core repulsion in the
slave-boson model. The parameter $m_r$ determines the coupling between
the bosons and the gauge field. The limit $m_r\to 0$ at constant
$\lambda_0$ corresponds to the vanishing of the perturbative
corrections to $\mathcal{L}^s$. We expect our results to be reasonable for 
small $m_r$.

For convenience, we use a rescaled frequency, $w=\omega
{m_\mathrm{f}/n_\mathrm{f}}$, and rescaled momentum
$u=pn^{-1/2}_\mathrm{f}$. We also define dimensionless propagators 
${\tilde D}_{w,u}=n_\mathrm{f} D_{\omega_n,q}/m_\mathrm{f}$, and 
${\tilde C}_{w,u}= n_\mathrm{f} C_{\omega_n,q}/m_\mathrm{f}^2$. 

The result is that:
\begin{equation}
\label{equ:K1_scaling_tilde}
(\delta \omega) \left( {m K^{(1)}_{\omega_n}} \right) =  
{(\delta w) m_r\over 2\pi} \int_0^\infty 
{ u \mathrm{d} u \over 2\pi} 
\left( {\tilde D}_{w,u}^s \right)^2
\end{equation}
and
\begin{eqnarray}
%\lefteqn{\hspace{-15pt} 
(\delta\omega)\Pi^{(1)}_{\omega_n} =
\frac{(\delta w) m_r^2}{\pi}  
%} \times }\nonumber\\ & &
\int_0^\infty\frac{u \mathrm{d} u}{2\pi}
{\tilde D}_{w,u}^s {\tilde C}_{w,u}
%{u^2 \over w^2 + m_r \alpha_w^2 u^2 ( \lambda_{0}^{-2}
%\beta_w U_r + \frac{1}{4} u^2 m_r)}
\label{equ:Pi1_scaling_tilde}
\end{eqnarray}
As mentioned above, we can see from these expressions that $m_r$ can be
regarded as a measure of the coupling strength: the effect of the gauge field
on the bosons is small if $m_r$ is small.
Our choice of renormalised variables means that $\tilde D^s$ 
is independent of $m_r$ and $U_r$. 

\begin{figure}[bht]
\epsfig{file=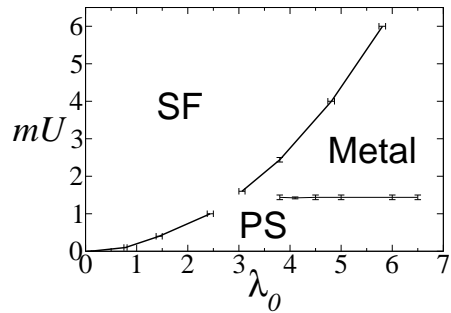, width=0.8\columnwidth}
\caption{Phase diagram for $m_r=1$. SF: superfluid; PS: phase separation;
Metal: vanishing superfluid response but finite compressibility.}
\label{fig:phs}
\end{figure}

Performing the scaling process numerically (at
$\Omega/E_\mathrm{F}=1$) gives a phase diagram as shown in
figure~\ref{fig:phs}. Our expansion (\ref{equ:L_rho_theta}) 
about the superfluid state in
means that these results are most reliable 
in the superfluid phase, up to the phase boundaries, which represent the
onset of the instabilities of the superfluid state. The metal/phase-separation
boundary in figure~\ref{fig:phs} is a rough guide only.

\begin{figure}[hbt]
\includegraphics{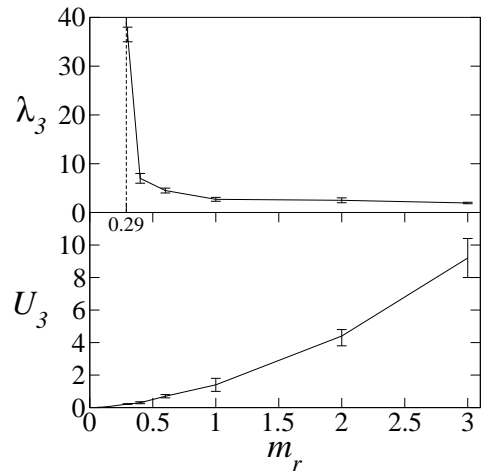} 
\caption{Position of critical point at varying $m_r$. $\kappa^{-1}$ and $n_s$
vanish together at the critical point $U=U_3$, $\lambda_0 =\lambda_3$. Metallic
region appears for $m_r>0.29$.
}
\label{fig:crit}
\end{figure}

We find that, at small $U_r$, there is an instability to phase
separation. This is because the correction to $\kappa^{-1}$, and hence
$U$, from (\ref{equ:K1_scaling}) is not directly dependent on $U$.  On
the other hand, at large $U_r$, we find that the superfluid is
stable. This is expected because $\Pi^{(1)}$ is small compared to
unity and $K^{(1)}$ is small compared to $U$, so that the corrections that we
calculate are small.

For intermediate values of $U_r$, we find that there may be a metallic
state: this is most likely at small doping (\emph{i.e.}, a
large bare penetration depth $\lambda_0$). However,
because $\Pi^{(1)}$ is a fractional correction and not an absolute
one, we note that it is not guaranteed that reducing the doping will
result in a metal even as $n\to 0$ ($\lambda_0\to\infty$).

The form of the phase diagram depends on $m_r$ (see
figure~\ref{fig:crit}). There is always a Meissner effect in the limit
$m_r\to 0$ (at constant $\lambda_0$). For small $m_r$, only the
instability to phase separation occurs. A metallic region appears in
the phase diagram at $m_r\simeq 0.3$.  Increasing $m_r$ above 0.3
favours both instabilities at the expense of the superfluid.

In terms of doping, the boson density decreases from left to right in
Fig.~\ref{fig:phs}. We see that the maximal doping for which the metal
exists is approximately $10\%$ for $m_r=1$ and $5\%$ for $m_r=5$,
provided that $U_r$ is not too small.

\subsection{Propagating gauge field}

We now compare the results for an overdamped gauge field with
the results for a propagating (Maxwell) gauge field, as used by
Feigelman~\emph{et~al.}~\cite{Ioffe}. The propagator is
$D_{\omega_n,q}=g^2 ( \omega_n^2+ c^2 q^2 )^{-1}$. As before, we rescale
the boson mass and the penetration depth, using the parameters in
the gauge field propagator. 
The relevant dimensionless parameters in this case are $U_r=m U$ as before; 
$\lambda_M^2 = (m g^2/ n)$ which is the rescaled penetration depth,
and $\alpha=g^2 / (mc^2)$: the coupling constant for the gauge field.
Rescaling frequency
and momentum via $w=(\omega_n / g^2)$, $u=cp/g^2$, 
the equations~\ref{equ:K1_scaling_tilde} and~\ref{equ:Pi1_scaling_tilde}
still apply, the only
changes being in the form of $\tilde D$, and in the replacement of
$m_r$ by $\alpha$ and $\lambda_0$ by $\lambda_M$.
In this case the momentum integrals in equations~\ref{equ:K1_scaling} 
and~\ref{equ:Pi1_scaling} can be done analytically, yielding: 
\begin{equation}
\left( m K^{(1)}_{\omega_n} \right) =
 {g^2\over 8\pi^2 mc^2} 
  {1\over g^{-2}\omega_n^2 + m^{-1} n_s(\omega_n)}
\end{equation}
\begin{eqnarray}
 \Pi^{(1)}_{\omega_n} &  = &
  {g^2\over 4 \pi^2 m c^2} {\kappa(\omega_n)\over n}
 {1\over \gamma(\gamma-1) + B} \times 
 \nonumber \\
& & \left[ 
\gamma \log \left( {\gamma^2 \over B} \right) + 
  {2B - \gamma \over X} \log\left( 
{ 1+X \over 1-X } \right) \right]
\end{eqnarray}
where $\gamma=\kappa(\omega_n)\left[\omega_n^2 + (n_s(\omega_n)g^2/m)\right]
/(4nmc^2)$,
$B=\omega_n^2 \kappa^2(\omega_n)/(4n n_s(\omega_n))$, and $X=\sqrt{1-4B}$ are
all dimensionless and are defined only for ease of writing.

\begin{figure}[hbt]
\epsfig{file=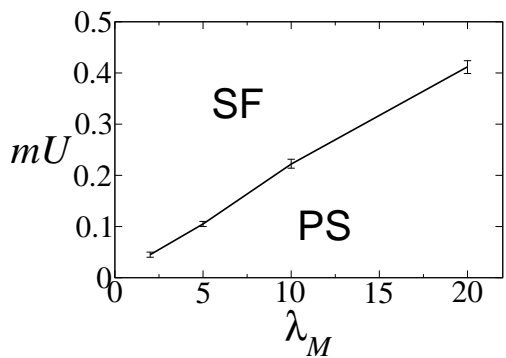, width=0.9\columnwidth}
\caption{Phase diagram for a Maxwell gauge 
  field with ${g^2/mc^2}=1$. SF: superfluid, PS: phase separation.}
\label{fig:max_phs}
\end{figure}

As discussed by
Feigelman~\emph{et~al.}~\cite{Ioffe}, the propagating gauge field combined 
with short-range repulsion tends to cause phase separation. This is
especially apparent at small $n_s$: the correction
$K^{(1)}_{\omega_n}$ diverges at small $n_s$ and $\omega$. 
This means
that the instability to phase separation will tend to dominate over the loss
of superfluid response, unless $n_s$ disappears at rather a high
frequency. The
phase diagram for moderate coupling strength, $\alpha=1$, is shown in 
figure~\ref{fig:max_phs}. The instability to phase separation is dominant at
this coupling, so there is no metallic phase.

\begin{figure}[hbt]
\epsfig{file=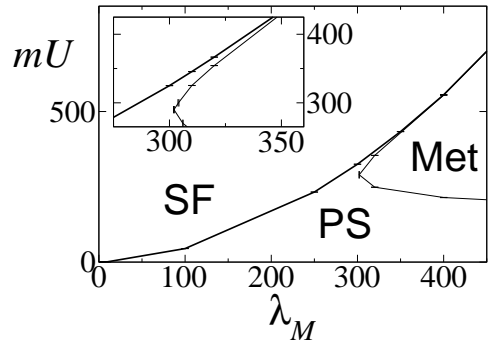, width=0.9\columnwidth}
\caption{Phase diagram for a Maxwell gauge 
field with ${g^2/mc^2}=3$, showing phase separated (PS), superfluid (SF) and
metallic (Met) regions. The inset shows the narrow phase separated region
between the metallic and superfluid regions. }
\label{fig:max_multi}
\end{figure}

At larger values of the coupling, around $\alpha\simeq 2$, a metallic phase
does appear; however, our one-loop perturbation theory is not reliable at these
couplings. A phase diagram is shown in figure~\ref{fig:max_multi}.
It is also clear from figure~\ref{fig:max_multi} that 
the metal requires very large values of 
the parameter $mU$ to stabilise the system against phase separation. So, in
contrast to the overdamped case, the parameters of the propagating field need
to be pushed to rather unphysical values to see a metallic phase.

Another consequence of the singularity in $K^{(1)}_{\omega_n}$ is that 
$\kappa$ is very strongly
renormalised near any metal-superfluid phase boundary. There is evidence
that $\kappa^{-1}$ will always vanish before $n_s$ in a very small region
near to the boundary: see figure~\ref{fig:max_multi}. 
The phase-separated region between
the metal and superfluid regions narrows with increasing $\lambda_M$. At
large values of $\lambda_M$, an apparent superfluid-metal transition does 
appear, but the numerical calculations show a discontinuity in the 
compressibility at the transition. This may be a symptom of an intervening
phase separated region, which is too narrow to resolved numerically.

As mentioned previously, the results of the renormalisation scheme are 
likely to be most valid in the superfluid phase, and on the boundaries of
that phase. The Maxwell case is distinct from the overdamped case in
that $K^{(1)}_{\omega_n}$ diverges as $n_s, \omega_n \to 0$, which has a 
strong effects on behaviour of the system very close to the phase boundary.

\section{Conclusion}

To summarise, we have analysed the effect of both overdamped and propagating
gauge fields on the ground state of a Bose liquid. In the overdamped case,
we find two different
instabilities of the superfluid phase. Specifically, the penetration
depth for the gauge field may diverge. This is interpreted as an
instability towards a homogeneous metallic phase. Alternatively, we
find that its compressibility may diverge. We interpret this to be
phase separation. We cannot comment on any spatial structures in the
new ground state because our treatment cannot identify any
characteristic length scale for this instability. A more careful
treatment of lattice effects and the compressibility at different
wavevectors is needed.

If the overdamped gauge field is replaced by a propagating one, then 
the instability to phase separation dominates over the divergence of
the penetration depth for all systems with local repulsion. This is in
agreement with Feigelman~\emph{et~al}~\cite{Ioffe}. 

Applying these results to the slave bosons that describe the charge
dynamics of the $t$--$J$ model, both instabilities are found at
dopings up to around 10\%. For $U_r>1$, a metallic phase is favoured
at these densities. (A large $U_r$ is natural for the hard-core slave
bosons.) Although precise numerical values may be affected by the
various approximations in our treatment, this Bose metal appears to be
in the regime in parameter space relevant to a
non-Fermi-liquid description of the normal state of the cuprates.

\begin{acknowledgments}
RLJ would like to acknowledge the support of a UK Engineering and
Physical Science Research Council studentship. DKKL is supported by a
Royal Society University Research Fellowship.
\end{acknowledgments}

\begin{appendix}

\section{Slave-boson decoupling of the $t$--$J$ model}
\label{sec:app-tj}

In this section we review the derivation of a bosonic model of the form given
in equation~\ref{equ:L_psi} from the $t$--$J$ model. Our notation follows
that of Lee and Nagaosa~\cite{LeeNag}.

The $t$--$J$ model is defined in equation~\ref{equ:tJ}.
We apply the `slave-boson' decoupling~\cite{Coleman,ReadNewns}
to this problem and write
$c_{i \sigma} = f_{i \sigma} b^\dagger_i$ where $f$ is a fermionic
spin-$\frac{1}{2}$ field and $b$ is a bosonic field. The bosons represent
charged `holes', and describe the charge dynamics of the system.
Neglecting terms quartic in the bosonic fields (valid at small doping)
we arrive at
\begin{eqnarray}
  H &  \simeq &  \sum_{\langle ij \rangle} \left[ -t \left(
  f^\dagger_{i\sigma} f_{j\sigma} b_i  b^\dagger_j + \hbox{h.c.} \right)
  - {J\over 2} f^\dagger_{i\sigma} f_{j\sigma}
               f^\dagger_{j\sigma'} f_{i\sigma'} \right]
 + \nonumber \\ & & i \sum_i \lambda_i
   \left( f_{i\sigma}^\dagger f_{i\sigma} + b_i^\dagger b_i -1 \right)
\end{eqnarray}
where the complex field $\lambda_i$ is a Lagrange multiplier enforcing the
constraint of no double occupancy.

Following Lee and Nagaosa~\cite{LeeNag} we make the mean field decoupling
$\sum_\sigma \langle f^\dagger_{i\sigma} f_{j\sigma} \rangle
= \chi_0 e^{i a_{ij}}$. We again neglect
terms quartic in the boson operators and we
arrive at the imaginary time Lagrangian:
\begin{eqnarray}
L_\tau & = & \sum_{i\sigma} f^\dagger_{i\sigma,\tau} \left[
     \partial_\tau - \mu_\mathrm{f} + i a^{0}_{i,\tau} \right] f_{i\sigma,\tau}
\nonumber  + \\
& &   \sum_{i} b^\dagger_{i,\tau} \left[
     \partial_\tau - \mu_b + i a^{0}_{i,\tau} \right] b_{i,\tau}
\nonumber - \\
  & &  {1\over 2} \chi_0 J \sum_{\langle ij \rangle, \sigma} \left[
      f^\dagger_{i\sigma,\tau} f_{j\sigma,\tau} e^{i a_{ij,\tau} } + \hbox{c.c
}
\right] \nonumber - \\
\label{equ:U1-latt}
& &   \chi_0 t \sum_{\langle ij \rangle} \left[
      b^\dagger_{i,\tau} b_{j,\tau} e^{i a_{ij,\tau} } + \hbox{c.c} \right]
\end{eqnarray}
where $a^0_{i}$ is the real part of $\lambda_i$.
The gauge invariance of the original electron operators under the
transformation
$f_{i\sigma} \to f_{i\sigma} e^{i \theta_i}$, $b_i \to b_i e^{i \theta_i}$ 
allows us to identify the $a_{ij}$ as the $U(1)$ gauge field associated with
this gauge symmetry.

If we assume that the relevant fields are slowly varying in space
and make the continuum approximation, we obtain a Lagrangian density
given by:

\begin{eqnarray}
\mathcal{L}_x  & = & \sum_\sigma f^\dagger_{\sigma,x} \left[
     \partial_\tau - \mu_\mathrm{f} + i a^{0}_x \right] f_{\sigma,x} +
\nonumber \\ & &
        b^\dagger_x \left[ \partial_\tau - \mu_b + i a^{0}_x \right] b_x +
\nonumber \\ & &
  {1\over 2 m_\mathrm{f}} \sum_\sigma
  f^\dagger_{\sigma,x} \left[ i \nabla - \bm{a}_x \right]^2 f_{\sigma,x} +
\nonumber \\ & &
  {1\over 2 m_b} b^\dagger_x     \left[ i \nabla - \bm{a}_x \right]^2 b_x
\label{equ:U1-ctm}
\end{eqnarray}
where the parameters $m_b$ and $m_\mathrm{f}$ are effective boson and
fermion masses respectively. They are related to the original $t$ and $J$,
and to $\chi_0$. The real 2--vector field $\bm{a}_x$ is defined by
$a_{ij,\tau} = (\bm{r}_i - \bm{r}_j)
  \cdot \bm{a}_{\frac{1}{2}(\bm{r}_i + \bm{r}_j),\tau}$.
In making the continuum approximation, we move from an original model with 
square symmetry to an isotropic model. 

The fermions
and bosons interact only via the gauge field, so integrating out the
fermions will give an effective propagator for the gauge field. The
effective gauge field action is, to quadratic order:

\begin{eqnarray}
S_g & = & {1\over 2 \beta L^2} \sum_{\omega_n,q}
   a_0^*(\omega_n,q) \Pi^0_\mathrm{ff}(\omega_n,q) a_0(\omega_n,q) + 
\nonumber \\
 & & {1\over 2 \beta L^2} \sum_{\omega_n,q}
     a_\perp^*(\omega_n,q) \Pi^\perp_\mathrm{ff}(\omega_n,q) a_\perp(\omega_n,q)
\end{eqnarray}
where we have fixed the gauge such that $\nabla \cdot \bm{a} = 0$;
only the transverse
component of $\bm{a}$ is non-zero and is denoted $a_\perp$.
The size of the system
is $L$ and the inverse temperature is $\beta$. The scalar part of the
gauge field has a propagator determined by
$\Pi^0_\mathrm{ff}(\omega_n,q)=
\langle \rho_\mathrm{f}^*(\omega,q) \rho_\mathrm{f}(\omega,q) \rangle$
where $\rho_\mathrm{f}(\omega,q)$ is the Fourier transform of the
fermion density.
At small energy and momentum $\Pi_\mathrm{ff}^0(\omega_n,q)$ 
tends to a constant, as predicted by
Thomas-Fermi screening: we model this field by a local replulsion $U$. This
is consistent with the introduction of the
scalar field to implement the constraint of exactly one fermion or one
boson per site.
The propagator of the vector part of $a$ is determined by
$\Pi_\mathrm{ff}^\perp(\omega_n,q) = T_{\mu\nu}
  \langle {J_\mathrm{f}^\mu}^*(\omega_n,q) J_\mathrm{f}^\nu(\omega_n,q) \rangle$ where
$J_\mathrm{f}^\mu(\omega_n,q)$ is the Fourier transform of the fermion current and
$T_{\mu\nu} = \delta_{\mu\nu} - {q_\mu q_\nu \over q^2}$ selects the
transverse part.

\section{Polarisation operator of an isotropic free Fermi gas}
\label{app:Pi}
In this section we discuss the overdamped gauge field propagator in a little
more detail. We first state the form of the free fermion polarisation that we
use and then discuss the effect of the cutoff $\Omega$.

The current-current correlation function of the free
fermions is given by
\begin{eqnarray}
\lefteqn{
\langle {J_\mathrm{f}^\mu}(\omega,q) J_\mathrm{f}^\nu(-\omega,-q) \rangle  =
  {2\over m_\mathrm{f}^2} \times
}  \nonumber \\   & &
\int {\mathrm{d} \omega' \over 2\pi}
{1\over L^2} \sum_{\omega',p}
          G^0_{\omega'+\omega,p+q} G^0_{\omega',p}
  (p+\frac{q}{2})^\mu (p+\frac{q}{2})^\nu
\label{equ:jf-jf}
\end{eqnarray}
where $G^0$ is the single-particle Green's function for a free
electron gas.

The imaginary part of this propagator is non-zero only in the particle-hole
continuum, where 
$q(q-2k_\mathrm{F}) < 2 m_\mathrm{f}|\omega | < q(q+2k_\mathrm{F})$. 
Within this region, we have:

\begin{eqnarray}
\lefteqn{
\mathrm{Im} \Pi^\perp_\mathrm{ff}(\omega,q)
 =
} \nonumber \\ & & {1\over 3 \pi m_\mathrm{f} q} \left[ \left( k^2 -
    \left( {m_\mathrm{f} |\omega| \over q} - {q\over 2} \right)^2 \right)^{3\over2}
   \right]_{k=k_\mathrm{min}}^{k_\mathrm{F}}
\end{eqnarray}
where $k_\mathrm{F}$ is the
Fermi momentum and
\begin{equation}
k_\mathrm{min} =
\mathrm{max}\left(\sqrt{k_\mathrm{F}^2-2m_\mathrm{f}|\omega |}, \left|{m_\mathrm{f}|\omega |\over q} -
  {q\over 2} \right| \right)
\end{equation}
The real part of the polarisation is given by:
\begin{eqnarray}
\mathrm{Re} \Pi^\perp_\mathrm{ff}(\omega,q)
 & = &
{12 m_\mathrm{f}^2 \omega^2 + q^4
\over 12 \pi m_\mathrm{f} q^2}
\end{eqnarray}
for $2 m_\mathrm{f}|\omega | < q(2 k_\mathrm{F} - q)$, and by:
\begin{eqnarray}
\mathrm{Re} \Pi^\perp_\mathrm{ff}(\omega,q)
 & = &
{12 m_\mathrm{f}^2 \omega^2 + q^4
\over 12 \pi m_\mathrm{f} q^2}
- \nonumber \\
& & {1\over 3 \pi m_\mathrm{f} q }
\left[ \left( {m_\mathrm{f} |\omega| \over q} - {q\over 2} \right)^2 -k_\mathrm{F}^2
\right]^{3\over 2}
\end{eqnarray}
in the rest of the particle-hole continuum.

\begin{figure}
\includegraphics{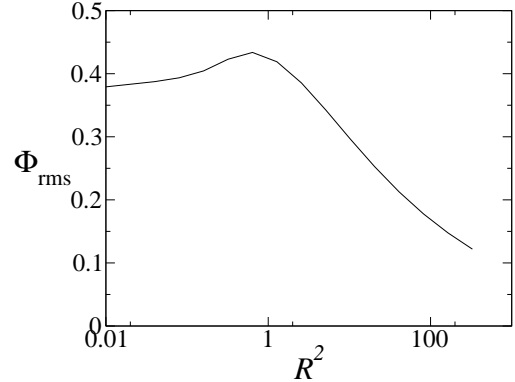}
\caption{Plot of root mean square flux through an area
$R^2$. Flux is measured in units of the flux quantum; area in
units of plaquette size. The fermion density is
one per plaquette.
Solid line is $\Omega=1000 E_\mathrm{F}$
(increasing $\Omega$ has no effect at these
areas. Dotted line is $\Omega=E_\mathrm{F}$.}
\label{fig:flux}
\end{figure}

As mentioned in section~\ref{sec:tJ}, we introduce an high-energy cutoff, 
$\Omega=E_\mathrm{F}$, to the spectral function of the gauge field propagator.
The effect of $\Omega$ on
the fluxes through various areas is shown in figure~\ref{fig:flux}. The
mean square flux through an area $R^2$ is calculated from:
\begin{eqnarray}
\langle \Phi^2 \rangle_R & = &
\int_{R^2} \mathrm{d}^2 \bm{r} \, \mathrm{d}^2 \bm{r}'
\langle (\nabla \times \bm{a})_{\bm{r},\tau}
        (\nabla \times \bm{a})_{\bm{r}',\tau} \rangle
\nonumber \\
& \simeq & R^4 \int_0^{\pi\over R} {p \mathrm{d} p \over 2\pi} p^2 D_{p,\tau=0
}
\end{eqnarray}

If no cutoff is used areas much smaller than one plaquette still
have large fluxes through them, which is not compatible with the 
underlying lattice in the
$t$--$J$ model; introducing the cutoff at the Fermi energy strongly
reduces fluxes through areas smaller than one plaquette, leaving the
behaviour at larger lengthscales unchanged.

\end{appendix}

\bibliography{my}
\bibliographystyle{apsrev}
\end{document}